\documentclass[aps,prx,reprint,superscriptaddress, floatfix]{revtex4-2}
\usepackage{amsmath,amssymb}
\usepackage{bbold}
\usepackage{float}
\usepackage{graphicx}
\usepackage{dcolumn}
\usepackage{bm}
\usepackage[colorlinks]{hyperref}
\usepackage{bbold}
\usepackage{graphicx}
\usepackage{dcolumn}
\usepackage[T1]{fontenc}
\usepackage{braket}
\usepackage{mathtools}
\usepackage{todonotes}
\usepackage{nicematrix}
\usepackage{textcomp, gensymb}
\usepackage{tabularx}
\usepackage{subcaption}
\usepackage{algorithm}
\usepackage{physics}
\usepackage{algpseudocode}
\usepackage{comment}
\usepackage{caption}
\begin{document}
\title{Optimization of chemical mixers design via tensor trains and quantum computing}

\author{Nikita~S.~Belokonev}
\address{Terra Quantum AG, Kornhausstrasse 25, 9000 St. Gallen, Switzerland}

\author{Artem~A.~Melnikov}
\address{Terra Quantum AG, Kornhausstrasse 25, 9000 St. Gallen, Switzerland}

\author{Maninadh~Podapaka}
\address{Evonik Industries AG, Rellinghauser Straße 1—11, 45128 Essen, Germany}

\author{Karan~Pinto}
\address{Terra Quantum AG, Kornhausstrasse 25, 9000 St. Gallen, Switzerland}

\author{Markus~Pflitsch}
\address{Terra Quantum AG, Kornhausstrasse 25, 9000 St. Gallen, Switzerland}

\author{Michael~R.~Perelshtein}
\email{mpe@terraquantum.swiss}
\address{Terra Quantum AG, Kornhausstrasse 25, 9000 St. Gallen, Switzerland}

\begin{abstract}
Chemical component design is a computationally challenging procedure that often entails iterative numerical modeling and authentic experimental testing. 
We demonstrate a novel optimization method, {\tt Te}nsor {\tt tra}in {\tt Opt}imization ({\tt TetraOpt}), for the shape optimization of components focusing on a Y-shaped mixer of fluids.
Due to its high parallelization and more extensive global search, {\tt TetraOpt} outperforms commonly used Bayesian optimization techniques in accuracy and runtime.
Besides, our approach can be used to solve general physical design problems and has linear complexity in the number of optimized parameters, which is highly relevant for complex chemical components.
Furthermore, we discuss the extension of this approach to quantum computing, which potentially yields a more efficient approach.
\end{abstract}

\maketitle

\section{Introduction}

The development of new reactors and components is opening up entirely new opportunities for the chemical industry to optimize the geometry of its plants and processes \cite{hoseini2021_shape_opt_reactors}. 
These opportunities include improving the plants’ performance, reducing costs and decreasing their environmental impact.
However, the design of various components (e.g., fluid  mixers) is a computationally challenging process \cite{norton2013cfd}. 

The underlying problem in designing these components is searching for the geometry (shape) that satisfies certain criteria. 
For instance, the maximization of the efficiency of a chemical process or the minimization of the mechanical tension in a component can be considered.
Such design is usually done in an iterative way, combining numerical modeling with real experimental testing \cite{choi2005investigation_Y_mixer}.
With complex systems, even numerical modeling takes a substantial amount of time and computational resources, so it is not feasible to perform many of these simulations.
The only efficient way to solve such a problem is by sampling the objective with various geometries and deciding what is the optimal set of geometrical parameters while attempting to keep the number of samples as low as possible.
In a given high dimensional parameter system with multi objective optimization, black-box optimization techniques can be very effective \cite{audet2017_balck_box}.

Black-box optimization deals with 
problems where the structure of the objective function and/or the constraints defining the set are unknown.
For example, in our case, there is no known analytical solution for the partial differential equations that govern the dynamics. 
The most naive way is to sample the objective using an equidistant grid in a parameter space. 
However, such an approach scales exponentially with the number of parameters and is clearly not feasible.
More advanced approaches include the use of Bayesian optimization \cite{tutorial_bayes_2010}, where the probability distribution is updated after each sample so the next sampling point is provided by the optimization routine.
While Bayesian optimization is widely used in academia and industry, its scaling is unclear and the structure of the algorithm does not allow for highly parallel processing or hardware acceleration, e.g. the use of Graphical Processing Units (GPUs).

In this work, we consider the tensor-train black-box optimization technique, {\tt TetraOpt} (see Ref.~\onlinecite{morozov2023} and \onlinecite{naumov2023}), and apply it to a specific component design, a Y-mixer used for the mixing of two fluids \cite{park2021_ymixer}.
We introduce a set of geometrical parameters and utilize {\tt TetraOpt} to find the geometry that provides the most efficient mixing of liquids.
We demonstrate that such an optimization method can be implemented in parallel so that it reduces the time of the component design.
This speeds up the prototyping and the full development cycle. 
In addition, this optimizer finds a better optimum in comparison to Bayesian optimization and requires much less work with hyperparameter tuning, since it has only three hyperparameters, which are very intuitive.

\section{Problem description}

The Y-mixer that is used to mix two liquids consists of two inlets that are connected symmetrically at a certain angle, guiding the liquids to a single outlet as shown in Fig.\,\ref{fig:Mixers}. 
In this work, we consider the mixing of water with ethanol with the property data given below.
\begin{table}[H]
	\centering
	\begin{tabular}{|l|l|l|}
		\hline
		  {\bf Parameter}               & {\bf Ethanol~~}      & {\bf Water~~}\\ \hline
        Density, $kg/m^3$       & 789               & 990   \\ \hline  
        Molar weight, $g/mol$   & 46                & 18     \\ \hline 
        Dynamic viscosity, $mPa \cdot s$~~~ & 1.18   & 1.0  \\ \hline 
	\end{tabular}
\end{table}
{\color{black} In order to simulate the flow of the liquids, we utilize an open-source Computational Fluid Dynamics (CFD) software package, {\tt OpenFoam} \cite{jasak2007openfoam}, using the {\tt PyMesh} utility for mesh generation \cite{zhou2016mesh}.
The simulation is performed using the {\tt reactingFoam} utility from {\tt OpenFoam}. 
The case is simulated using a {\tt kEpsilon} turbulence model \cite{jasak2007openfoam}.
The volumetric flow rate is fixed for both liquids at $8$ mL/s.}

 \begin{figure*}
    \noindent\centering{
    \includegraphics[width = 0.44\linewidth]{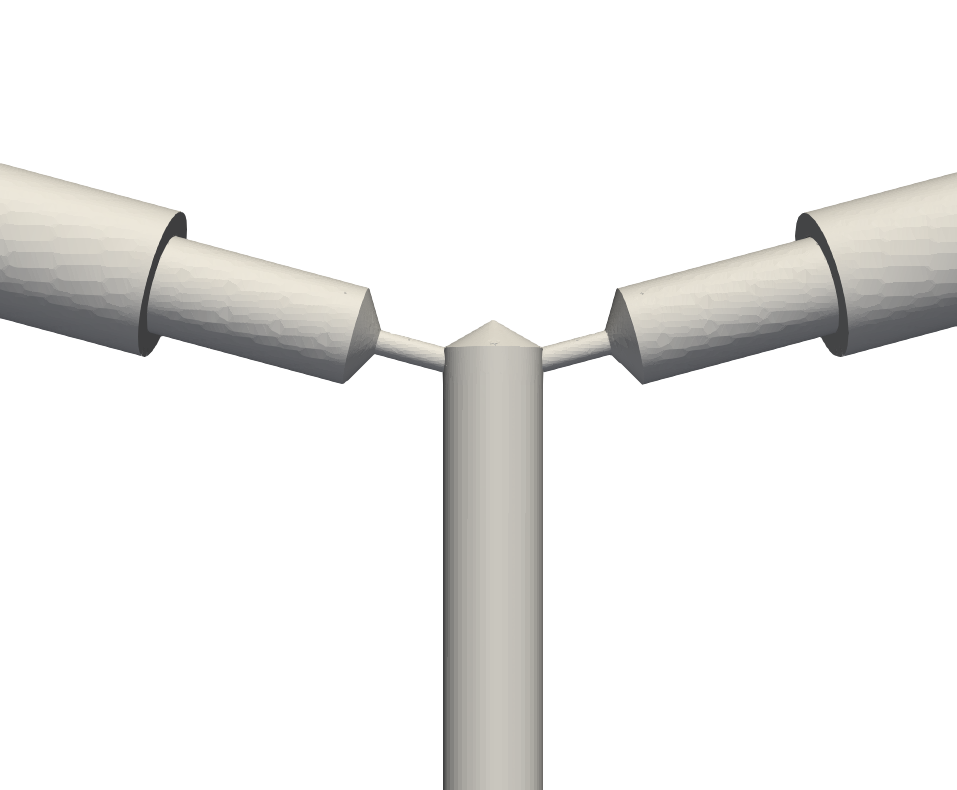} \qquad \qquad
    \includegraphics[width = 0.4\linewidth]{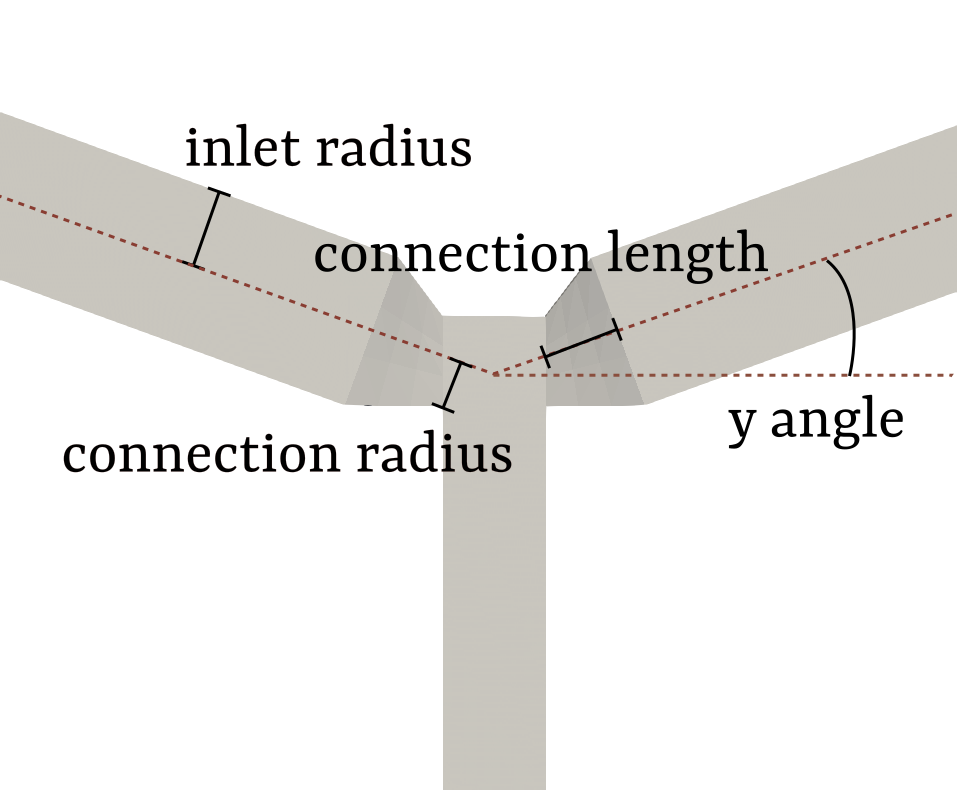}}
    \caption{{\color{black} Left: The original Y-mixer considered for the mixing of two fluids.
    Right: The simplified geometry of the Y-mixer With the parameters considered for the optimization.}}
    \label{fig:Mixers}
\end{figure*}

The shape of a Y-mixer includes a round section and different diameters along the tubes, as shown in Fig.\,\ref{fig:Mixers} (left). 
The inlet tubes have three different diameters along their lengths, while the outlet tube has a constant diameter.
When a Y mixer with narrow channels and a long outlet tube is considered, a detailed simulation with a fine mesh (a few million elements) could take a few hours until convergence.
In order to benchmark optimization methods, we simplify the shape, as shown in Fig.\,\ref{fig:Mixers} (right), which reduces the runtime to 20 seconds.
Such a simplification affects both the dynamics inside the mixer and the objective function but without loss of generality, we can use it with the key goal to analyze and benchmark the {\tt TetraOpt}.

In order to numerically characterize the quality of the mixing, we calculate the coefficient of variation (CoV) of the phase fraction of the liquids at a horizontal section, \textcolor{black}{which is $2.5$\,mm below the mixing chamber.}
The CoV is calculated in the following way:
\begin{equation}
    CoV = \frac{\sigma(m_1/(m_1+m_2))}{\langle m_1/(m_1+m_2) \rangle},
\end{equation}
where $m_1(x)$, $m_2(x)$ are phase fractions of the corresponding liquids on the section surface, which we obtain by solving the Navier-Stokes equations with {\tt OpenFoam}.
The standard deviation of $x$ is $\sigma(x)$ and the mean value of $x$ is $\langle x \rangle$. 
For homogeneous mixing, the coefficient of variation is close to zero ($CoV \rightarrow 0$), which constitutes the optimization problem that we solve here.

Here, we consider a set of four parameters to optimize:
\begin{enumerate}
    \item {\bf y-angle}, the angle between the inlet tube and the outlet tube (from $0^\circ$ to $30^\circ$);
    \item {\bf connection radius}, the effective radius of the closest to the outlet part of the inlet tube (from 0.2 to 0.5 mm);
    \item {\bf connection length}, the length of the closest to the outlet part of the inlet tube (from 0.5 to 1.5 mm);
    \item {\bf inlet radius}, the radius of the inlet (from 0.2 to 0.6 mm).
\end{enumerate}

{\color{black} Due to the nature of the problem, it is impossible to write down the CoV as a function of these four parameters since it involves the solution of the Navier-Stokes equation.
Therefore, the problem is considered to be a black-box -- it is possible to sample the CoV at arbitrary parameters values with the goal of finding the minima of the CoV as fast as possible.
}

\section{Bayesian optimization}

One of the most common ways to solve black-box optimization problems is to use Bayesian optimization \cite{frazier2018tutorial_bayessian}. It is successfully used in hyperparameter tuning tasks in machine learning \cite{wu2019hyperparameter, sagingalieva2022hyperparameter} and in shape optimization in CFD problems \cite{morita2022_bayes_shape}. 
The only assumption we make is that the cost function is continuous and its value can be estimated at a given point in a specified search area, which is true for the given shape optimization problem.

The Bayesian algorithm works in an iterative way -- it leverages obtained information about the function (such as the function values at several points) to approximate it.
At each iteration, it provides a new point at which the function should be estimated and updates the approximation using the new point.
The process is repeated until it converges.

\begin{figure}[ht]
    \noindent\centering{
    \includegraphics[width = 1 \linewidth]{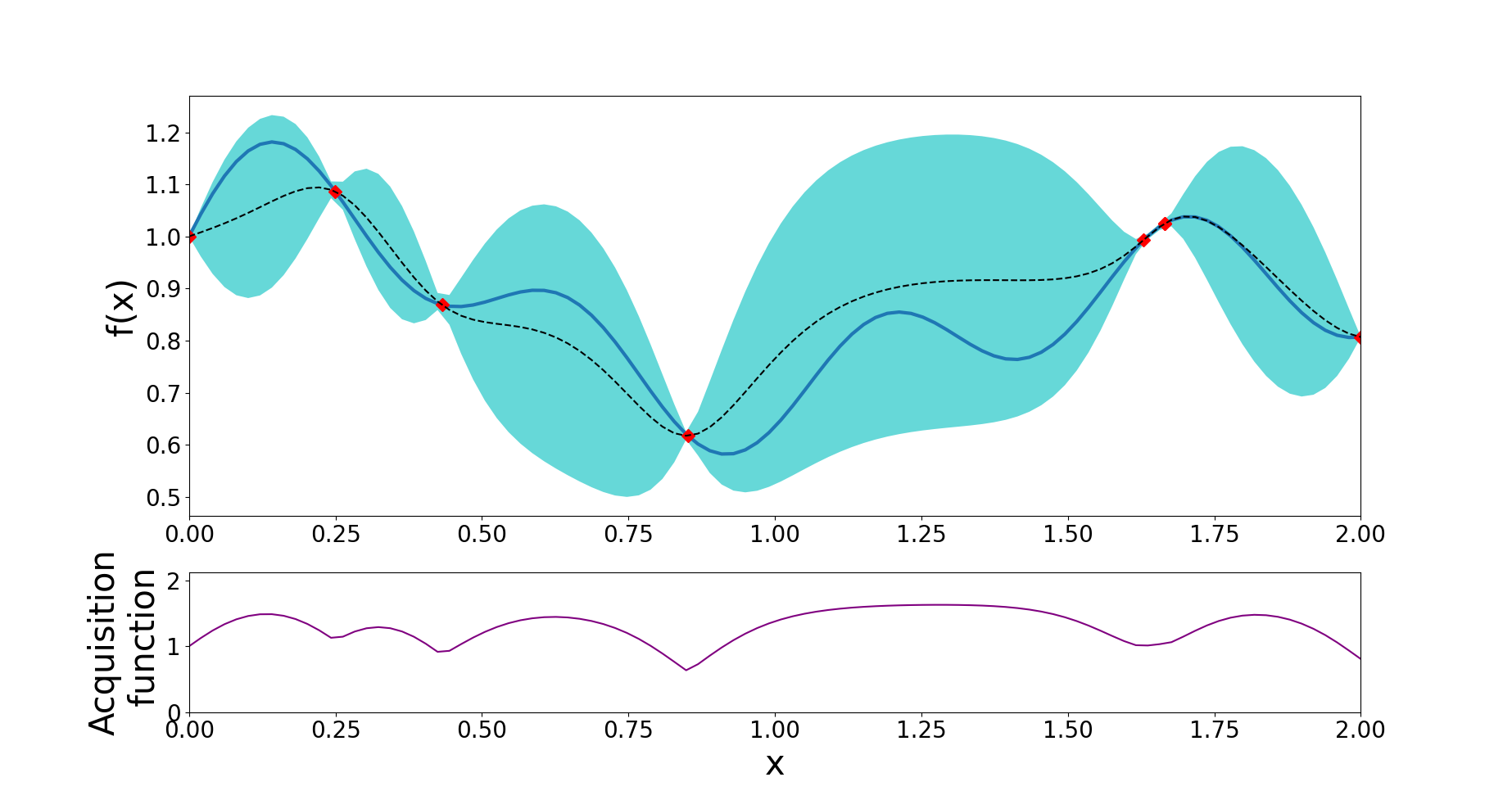}}
    \caption{Bayesian optimization. 
    The blue curve represents the target function, the red dots depict the evaluated points, the dashed line shows a mean $m(x)$ function and the blue region covers $[m(x) - \sigma(x), m(x) + \sigma(x)]$ area. 
    The acquisition function is plotted below.}
    \label{gaussian_process}
\end{figure}

\begin{figure*}
    \includegraphics[scale = 0.4]{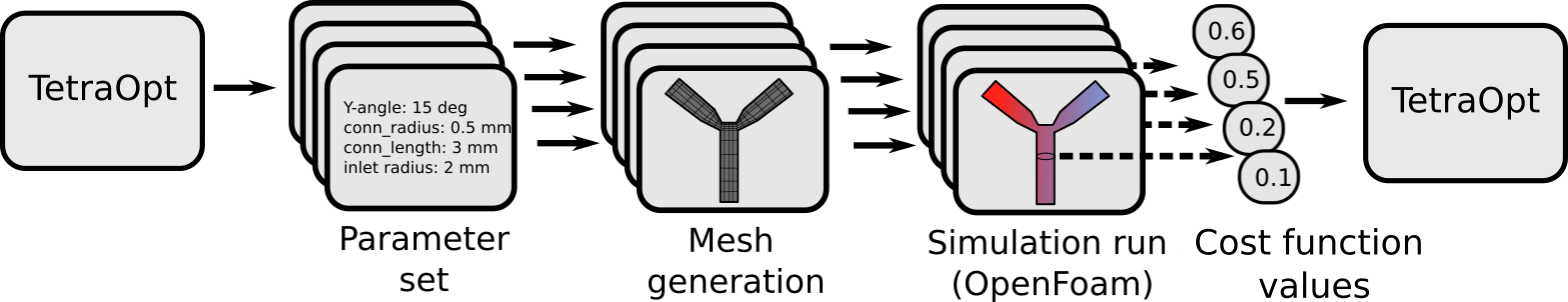}
    \caption{Overall scheme of shape-optimization via TetraOpt. 
    Firstly, {\tt TetraOpt} generates a list of geometrical parameter sets at which the simulation should be performed. 
    Then, the meshes for corresponding geometries are generated in {\tt Python} using the {\tt PyMesh} utility. 
    After that, at each set of parameters, the simulations are run in parallel utilizing the previously generated meshes and {\tt OpenFoam} (this is the most computationally difficult step). 
    The cost values are calculated from each simulation using Python and inner {\tt OpenFoam} functions. 
    Then the data is passed to {\tt TetraOpt} and the cycle is repeated until convergence.}
    \label{fig:overall_scheme}
\end{figure*}

In the beginning, the algorithm receives the initial function values at several points.
A single iteration of the algorithm consists of three steps:
\begin{enumerate}
    \item Based on all the available data about previously estimated points, an approximation is built (a surrogate model), which is usually done via Gaussian Processes \cite{wu2019hyperparameter}. 
    A Gaussian process (GP) is defined by its mean function $m: \mathbf{x} \rightarrow \mathbb{R}$ and its covariance function $k: \mathbf{x} \times \mathbf{x} \rightarrow \mathbb{R}$, which we denote as
$$
f(\mathbf{x}) \sim \mathrm{GP}\left(m(\mathbf{x}), k\left(\mathbf{x}, \mathbf{x}^{\prime}\right) \right).
$$
    In the one-dimensional case, it finds the mean $m(x)$ and the dispersion $\sigma(x)$ functions as shown in Fig.\,\ref{gaussian_process} (top).

    \item The second step is to build the acquisition function, which shows how likely a point should be chosen as the next estimate in terms of the exploration-exploitation trade-off. Exploitation means sampling at points where the surrogate model predicts a high objective and exploration means sampling at locations where the prediction uncertainty is high. 
    The simplest acquisition function in the case of maximization is the following:
    
    \[u\left(\mathbf{x} \mid \mathcal{D}_{1: t-1}\right) = m(\mathbf{x}) + \sigma(\mathbf{x}),\]

    which corresponds to the upper blue covering line in Fig.\,\ref{gaussian_process} (bottom). 
    However, the more popular ones are the maximum probability of improvement (MPI), expected improvement (EI) and upper confidence bound (UCB) \cite{tutorial_bayes_2010}. Here $\mathcal{D}_{1: t-1}=\left(\mathbf{x}_1, y_1\right), \ldots,\left(\mathbf{x}_{t-1}, y_{t-1}\right)$ are the $t-1$ samples drawn from $f$ so far.
    \item The next sampling point $\mathbf{x}_t$ is determined 
    according to
    $\mathbf{x}_t=\operatorname{argmax}_{\mathbf{x}} u\left(\mathbf{x} \mid \mathcal{D}_{1: t-1}\right)$. Then the function value at this point is obtained as $y_t=f\left(\mathbf{x}_t\right)$. Finally,
    the sample is added to previous samples $\mathcal{D}_{1: t}=\mathcal{D}_{1: t-1},\left(\mathbf{x}_t, y_t\right)$ and the GP is updated.
\end{enumerate}

The main disadvantages of the algorithm are that it is poorly parallelizable and struggles to work with non-continuous variables. 
The poor parallelization is due to the fact that the algorithm works sequentially and at each step of the optimizer's work, it is necessary to estimate the value of the cost function at only a single point.
Even though multiple runs of the algorithm can be performed in parallel, it usually fails to significantly change the efficiency \cite{frazier2018tutorial_bayessian}.

\section{Tensor Train optimization}\label{TetraOpt}
Here, we propose to use a completely different black-box optimization algorithm, which is based on Tensor Train (TT) decompositions \cite{sozykin2022ttopt, zheltkov2020global}. 
Tensor Trains \cite{TT_main_paper} represent multi-dimensional tensors in a compressed form as a product of small tensors:

\begin{align*}
    A\left(i_{1}, \ldots, i_{d}\right)= \sum_{\alpha_{0}, \ldots, \alpha_{d-1}, \alpha_{d}} & G_{1}\left(\alpha_{0}, i_{1}, \alpha_{1}\right) G_{2}\left(\alpha_{1}, i_{2}, \alpha_{2}\right) \\ & \ldots G_{d}\left(\alpha_{d-1}, i_{d}, \alpha_{d}\right),
\end{align*}    
where $G_j$ is the 3-dimensional tensor called the TT-core.
The indices $i_j$ run through values from $1$ to $n$. 
The main characteristic of such a representation is the rank, $r$, which is equal to the maximum size among the indices $\alpha_{0}, \alpha_{1}, \ldots, \alpha_{d}$ and which expresses the correlations between variables/indices.

As in a grid search, the optimization algorithm ({\tt TetraOpt}) requires discretization of the search space on a uniform grid. 
Let $d$ be the number of variables and $n$ be the grid size in one dimension. 
However, unlike grid search, {\tt TetraOpt} does not estimate the cost function at all $n^d$ points of the grid but instead dynamically provides the next set of evaluating points in the search space based on the knowledge accumulated during all previous evaluations, as in Bayesian optimization.
The main advantage of this algorithm in comparison to Bayesian optimization is its ability to be run in a parallel way and provide a better (``more global'') search for the optimum.

The main hyperparameters of the {\tt TetraOpt} algorithm are the \textit{number of variables}, $d$, the \textit{grid-size} in one dimension, $n$, and the \textit{rank}, $r$, with which we try to approximate a tensor of discretized function values via tensor train decomposition. 
The larger $r$ is, the better the optimum is but this requires more time. 
The last hyperparameter is the \textit{number of iterations}, $I$. 
{\tt TetraOpt} requires $O(I dnr^2)$ function calls and performs $O(Idnr^3)$ calculation operations. 
Since black-box functions are usually hard to estimate, the runtime of the algorithm can be neglected in comparison to the time of the function calls.

{\tt TetraOpt} is built upon the cross-approximation technique ({\tt TT-cross}) \cite{tt_cross}, which enables the approximation of a given tensor in the Tensor Train format by referencing only a subset of its elements.
The {\tt TT-cross} algorithm in turn is based on the {\tt MaxVol} routine \cite{maxvol}, which finds an $r \times r$ submatrix of maximum volume (i.e. a submatrix with a maximum determinant module) in an $n \times r$ matrix. 
It can be shown that the maximum element of a submatrix with maximal volume is a good approximation of the maximal element of the whole matrix \cite{maxvol}:
\[\hat{J}_{\max } \cdot r^2 \geq J_{\max },\]
\textcolor{black}{
in terms of its modulus, $\hat{J}_{\max}$ is the maximal element of a $r \times r$ submatrix with maximal volume, and $J_{\max }$ is the maximal element of the whole matrix.}
\textcolor{black}{To get more intuition and learn the technical aspects of tensor-based algorithms for optimization, we refer the reader to Ref.\,\onlinecite{sozykin2022ttopt}.}

\begin{figure}
\centering
\includegraphics[width=1\linewidth]{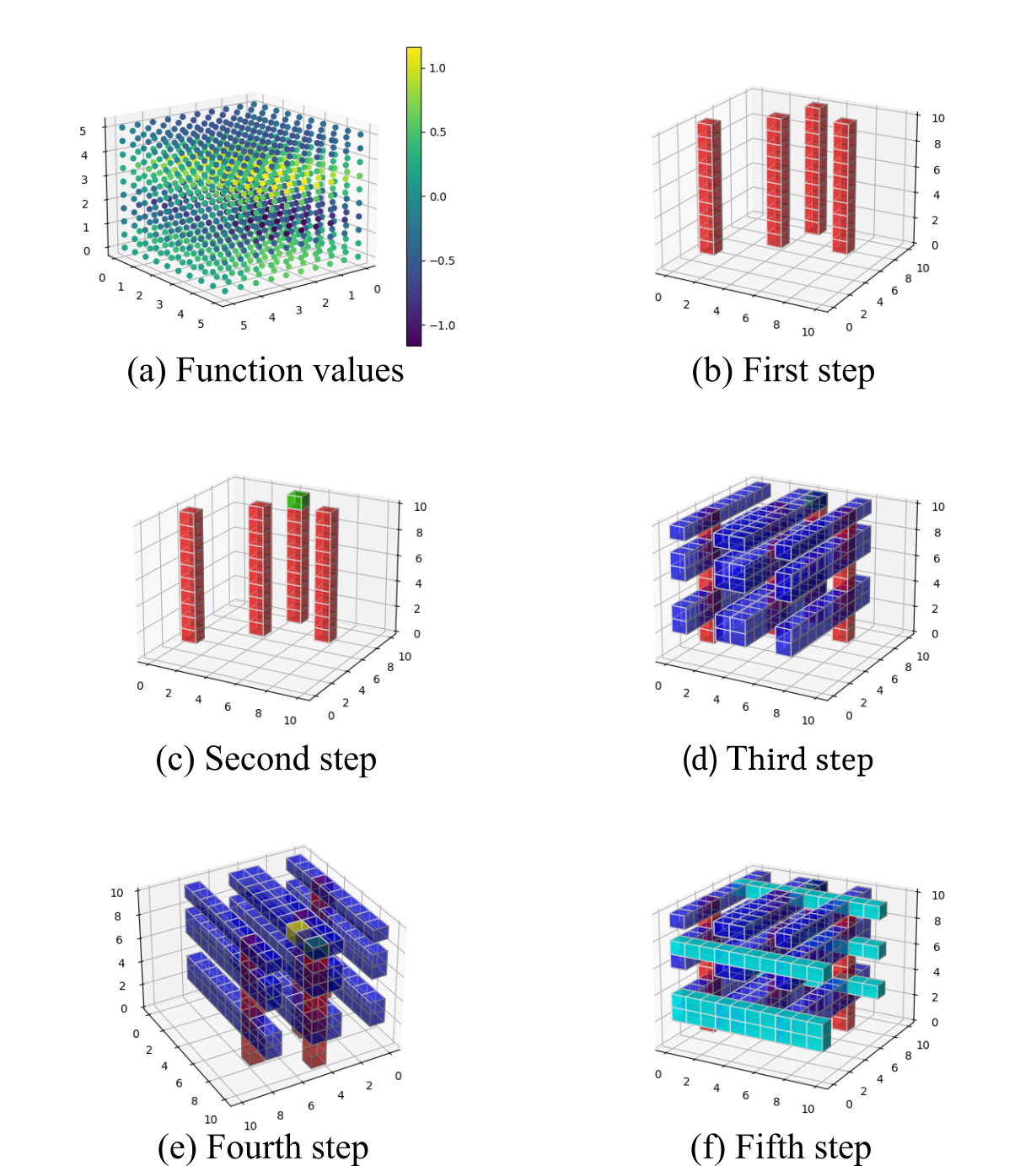}
\caption{The sequence of steps in {\tt TetraOpt} using a 3D function example. 
(a) Demonstration of the function discretization on a uniform 3D-grid; the figures
(b)-(f) indicate the points in which the function is estimated during one iteration of the {\tt TetraOpt} algorithm. 
The next step is shown in a bright color and the previous steps are transparent.}
\label{fig:TetraOpt}
\end{figure}

The overall scheme of the {\tt TetraOpt} workflow can be found in Fig.\,\ref{fig:overall_scheme}: {\tt TetraOpt} iteratively requires estimating the values of the optimization function at several grid points (see Fig.\,\ref{fig:TetraOpt}), which we count using the {\tt OpenFoam} simulation.
In terms of finding the optimum, the algorithm remembers the best points estimated during the {\tt TT-cross} algorithm and updates them if superior points are found.
In addition, the algorithm is parallelizable because,
at each step, it requires an estimation of the cost function at $nr^2$ points, which can be done in parallel. 
However, there is no rigorous proof that it finds a better optimum than Bayesian optimization - each task requires a separate analysis. 

\section{Results}
\begin{figure}[h!]
    \centering
\includegraphics[width=1\linewidth]{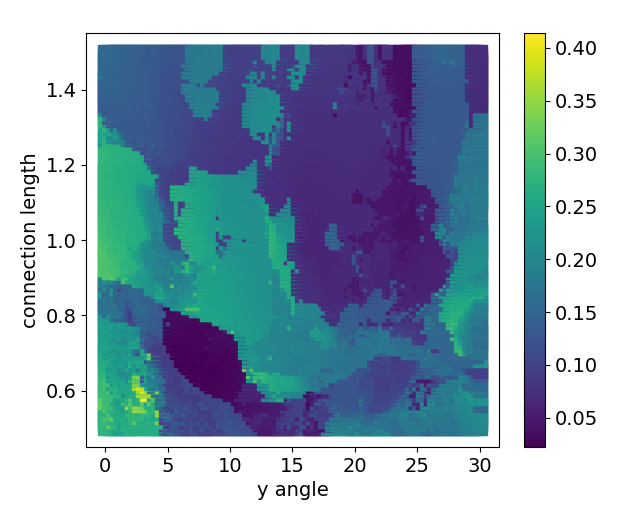}
    \caption{The cost function landscape as a function of two parameters: connection length and y-angle. 
    The darker the colour, the better the mixing is. 
    The fixed coordinates are the inlet radius = 0.275~mm and the connection radius = 0.3~mm. 
    We sample 100 points on each axis.}
    \label{fig:shapeopt_landscape}
\end{figure}

{\color{black}
In this section, the results of the shape optimization using the Tensor Train optimization technique and its comparison with Bayesian optimization are presented.
The grid-search was utilized to find a global optima but, of course, such a method is computationally inefficient.
All computations are performed using the {\tt QMware} system \cite{white_paper_tq}, including CFD simulations and optimization.
The CoV dependency on the given variables is complicated and non-obvious near the optimal value, as can be seen in Fig.\,\ref{fig:shapeopt_landscape}.}

\begin{figure}
    \includegraphics[width=1\linewidth]{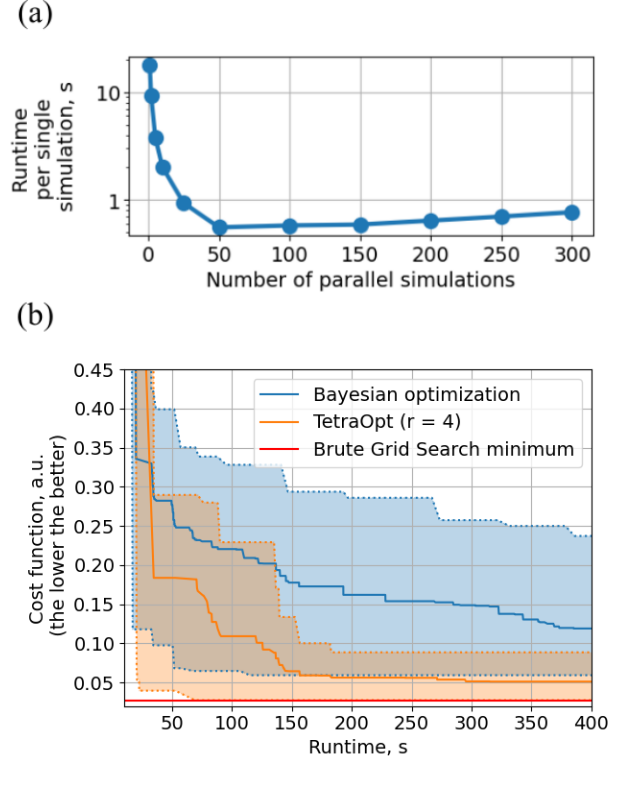}
    \caption{
    (a) Execution time per simulation dependent on the number of parallel simulations.
    The effective time of each simulation time decreases but after 50 parallel simulations, it does not change (or even increases) due to imperfect parallelization methods and hardware.
    (b) Comparison between {\tt TetraOpt} and Bayesian optimization. 
    {\tt TetraOpt} finds, on average, an approximately $2.33$ times better optimum in the same time than the Bayesian approach.
    Here, {\tt TetraOpt} performs CFD simulations in parallel, while Bayesian optimization requires only running one simulation at each iteration.
    Shadowed areas denote the variations in the behavior of the optimizers during 10 launches: dotted lines represent the best and worst scenarios for each optimizer.
    The red line represents the minimum found by a grid search.}
    \label{fig:shapeopt_comp}
\end{figure}

Based on the simplified CFD model, a single simulation of {\tt OpenFOAM} takes 16 seconds.
As was stated, the {\tt TetraOpt} algorithm is highly parallelizable, thus we run multiple CFD simulations in parallel so the effective runtime was reduced to $1.1$ seconds per simulation, 
as shown in Fig.\,\ref{fig:shapeopt_comp}(a).

The results of the comparison between {\tt TetraOpt} and the Bayesian optimization are shown in Fig.\,\ref{fig:shapeopt_comp}(b), 
\textcolor{black}{where we average the results by running both algorithms 10 times (solid curves).
We compare two optimizers in terms of the runtime and demonstrate that {\tt TetraOpt} converges faster to a more optimal value --} on average {\tt TetraOpt} obtains an approximately $2.33$ times better optimum at the end of optimization than Bayesian optimization: $0.051$ vs $0.12$.
Moreover, the tensor-based optimizer finds the best possible shape with a $0.027$ cost function value (0\% gap), while the Bayesian approach is able to find the shape with a $0.059$ cost value ($\sim$ 118\% gap).

 \textcolor{black}{
The behavior of the curves in Fig.\,\ref{fig:shapeopt_comp}(b) has a stepwise character due to the fact that CFD simulations are iteratively performed, which takes a considerable time (plateaus), and the optimum is updated while simulations are done (drops).}
The sharp jumps of the {\tt TetraOpt} curve (e.g., at $75$ and $125$ seconds) are due to the fact that a large number of points were simultaneously estimated, which significantly updated the optimum.

The found optima are shown in Table~\ref{table:results_comparison}. 
Remarkably, despite the fact that {\tt TetraOpt} requires an order of magnitude more function calls (number of carried-out simulations), it finishes the optimization in less time due to parallelization.
It is worth noting that the grid was such that the minimal value on the grid was close to the minimal value on a whole domain because otherwise, the Bayesian optimization would have some privileges.

\begin{figure*}
\centering
  \includegraphics[width=1\linewidth]{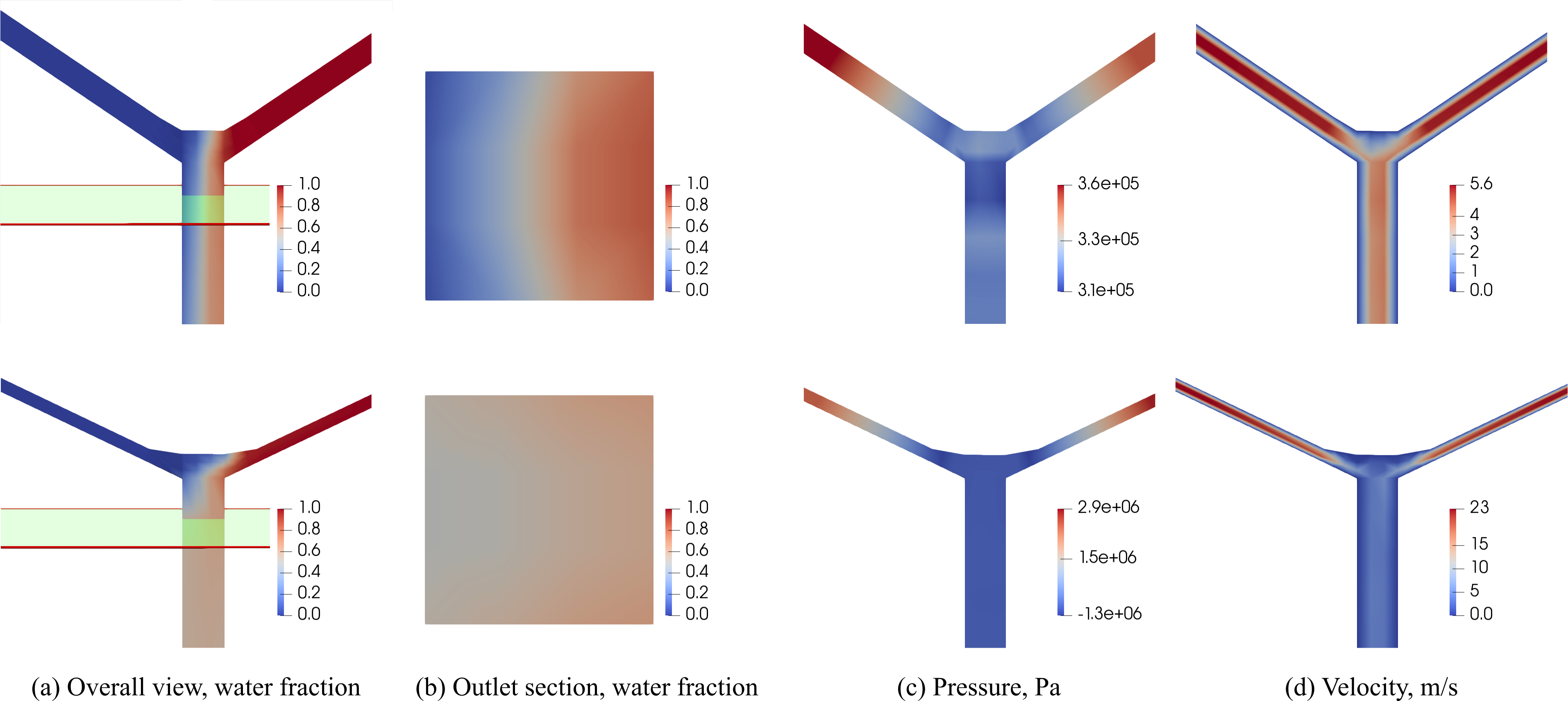}
\caption{The top row shows the data before optimization and the bottom row displays the data after optimization. 
(a) The water fraction with the cutting plane where the cost function is calculated (in green). 
(b) The water fraction at the specified cutting plane.
In the non-optimized case, the value of cost function (CoV) is 0.56, while after the optimization it is 0.05. 
The color represents the phase fraction of the water at each point. 
(c) The pressure before and after the optimization and (d) velocity distribution before and after the optimization.}
\label{fig:optres}
\end{figure*}

{\color{black}

In order to maximize the performance of both algorithms, we tune the hyperparameters. 
The following values are used for {\tt TetraOpt} and Bayesian optimization, respectively:
\begin{table}[H]
	\centering
	\begin{tabular}{l l|l l}
		\hline
		  {\bf TetraOpt}               & ~ & {\bf Bayesian}                   & ~\\ 
        \hline
	\# of iterations                     & 2~          &  \# of iterations  & 30 \\ \hline
Rank                      & 4          &   \# of initial points~~~~~   & 5 \\ \hline
Grid size per parameter~~~~~                   & 5          &   Acquisition method & UCB\\ \hline
	\end{tabular}
\end{table}
For the Bayesian optimization, a parameter $\kappa$, which indicates how close the next parameters are sampled, is $2.576$, which is constant on each iteration.
The Bayesian optimizer is taken from an open-source library \cite{beyes_library}. }

{\color{black}
Fig.\,\ref{fig:optres} shows a visual comparison between the flow behaviors in optimized and non-optimized geometries. 
Figs.~\ref{fig:optres}(a) and \ref{fig:optres}(b) represent the water fraction: while frame (a) shows overall profile of water fraction along the walls, frame (b) shows the water fraction distribution on the outlet section where the cost function is evaluated. 
It is clear that the water fraction in the optimized mixer case is more uniform and the absolute value is significantly smaller on average in comparison to the non-optimized mixer case.
Figs.~\ref{fig:optres}(c) and \ref{fig:optres}(d) show the pressure and absolute magnitudes of the velocity.}
\begin{table}[H]
\centering
	\begin{tabular}{|l|l|l|l|}
		\hline
  {\bf №}  	 &	  {\bf Parameter}         & {\bf TetraOpt}  & {\bf BayesOpt}   \\ \hline
      1 	 &      Y-angle, angle        & 26.1$^\circ$          & 7.5$^\circ$              \\ \hline
        2 	 &      Inlet radius, mm        & 0.27           & 0.275             \\ \hline
        3 	 &      Connection length, mm~   & 0.85          & 0.75              \\ \hline
        4 	 &      Connection radius, mm   & 0.28           & 0.3       
        \\ \hline
	\end{tabular}
~\\[7pt]
 	\begin{tabular}{|l|l|l|}
		\hline
    {\bf Resulting values}         & {\bf TetraOpt}  & {\bf BayesOpt}   \\ \hline
           Cost function, a.u.  ~~~~~~~~~~~   & 0.0274         & 0.0593             \\ \hline
              Average fun. calls          & 228.5        & 35                \\ \hline
          	     Runtime, s              & 325           & 440               \\ \hline
	\end{tabular}
    \caption{(1-4) Optimal parameters of the Y-mixer found by {\tt TetraOpt} and Bayesian optimization. 
    We show the final cost function value, 
    the total number of cost function calls (number of performed simulations) averaged by $10$ optimization runs and the total runtime of the optimization launched using {\tt QMware}.}
    \label{table:results_comparison}
\end{table}


\section{Extension to Quantum Computing}
Remarkably, a slightly modified version of {\tt TetraOpt} can be improved with a quantum algorithmic part. 
As stated in the description of the optimizer (Sec.~\ref{TetraOpt}), the algorithm tries to approximate the tensor of function values on a grid but it obtains the optimum only as a by-product of the approximation algorithm.
Therefore, the next step is to use the approximation to obtain new, even better, optima.

Let us denote a tensor of cost-function values on the uniform grid as $x$ (for later convenience, it is worth mentioning that we can always reshape $x$ into a vector), which we try to approximate via Tensor Train $x_{TT}$ of rank $r$ using the {\tt TT-cross} algorithm. 
The optimization task can now be redefined as the problem of finding the maximum (minimum) element of the tensor $x$. 
Here, we assume that $x_{TT}$ approximates $x$ with good precision, or at least its optimal values are close to the optima of $x$.

Thus, we can implement the power method \cite{chertkov2022optimization_power}. 
That is, we find the maximal element of $x^n$ instead of $x$, which is much easier since the largest element in $x^n$ is much larger, in relative terms, compared to the largest element of $x$. 
Since we assume that the optima of $x_{TT}$ are close to the optima of $x$ and since it is much faster to operate with $x_{TT}$ (for example, the squaring operation $x^2_{TT}$ costs $O(dnr^4)$ in TT format as compared to the complexity $O(n^{d})$ for classical squaring $x^2$), the idea is to realize the power method via Tensor Trains. 
However, the problem with this approach is that the ranks increase dramatically as $r^n$, leading to a significant increase in complexity. 
As a result, implementing this algorithm on a classical computer may not be as efficient as on a quantum computer, which does not have any complexity dependence on the ranks.
To use a quantum computer for this purpose, we only need efficient preparation of $x_{TT}$ and efficient multiplication by $x_{TT}$. Fortunately, several algorithms exist for encoding Tensor Trains into a quantum computer \cite{MPS_preparation, two_qubits_MPS_encoding, auto_dif_MPS_preparation, rudolph2022_shallow_decomposition}. Thus, a quantum implementation of this algorithm will be able to perform optimization in a more efficient way.\\

\section{Conclusion}

In this work, we utilized {\tt TetraOpt} to solve a shape optimization problem of a Y-mixer used for the mixing of two fluids. 
This problem is considered a black-box optimization, i.e. there is no explicit expression for the cost function and its estimation at a single point requires significant computational resources.
We demonstrated that compared to Bayesian optimization, {\tt TetraOpt} finds a much better optimum in less time. 
We concluded that such an improvement comes from the fact that {\tt TetraOpt} is a parallel technique and performs better exploration during the optimization compared to Bayesian methods.

It is worth emphasizing that the application of this method is not limited to the task at hand but can be applied to any optimization problem since it requires only an objective function given at a point in the search area. 
Besides, the method is straightforward to use because there are only three hyperparameters with intuitive settings.

Furthermore, we demonstrate an extension of this method to quantum hardware -- the realization of the power method via quantum circuits, which can provide a better optimum. 
The implementation of the quantum part and its application to more complex problems and geometries is the subject of future work.

\bibliography{lib}
\bibliographystyle{unsrt}

\end{document}